\pdfoutput=1
\documentclass[usenatbib]{mn2e}
\usepackage{apjfonts}
\usepackage{amssymb}
\usepackage{amsmath}
\usepackage{ctable}
\usepackage{url}
\usepackage{fixltx2e} 

\newcommand{\be}{\begin{equation}}
\newcommand{\ee}{\end{equation}}

\newcommand{\msun}{M_{\sun}}

\newcommand\plotone[1]
 {\centering \leavevmode \includegraphics[width={0.99\columnwidth}]{#1}}

\newcommand{\acknowledgments}{\begin{small}\section*{Acknowledgments}\end{small}}
\newcommand\altaffilmark[1]{$^{#1}$}
\newcommand\altaffiltext[1]{$^{#1}$}
\title[AGN-Host Galaxy Misalignments]{Why Are AGN and Host Galaxies Misaligned?\vspace{-0.5cm}}

\author[Hopkins et al.]{
\parbox[t]{\textwidth}{ 
Philip F. Hopkins\altaffilmark{1}\thanks{E-mail:phopkins@astro.berkeley.edu}, 
Lars Hernquist\altaffilmark{2}, 
Christopher C.~Hayward\altaffilmark{2}, 
Desika Narayanan\altaffilmark{3}
} 
\vspace*{6pt} \\
\altaffiltext{1}{Department of Astronomy, University of California
  Berkeley, Berkeley, CA 94720} \\
\altaffiltext{2}{Harvard-Smithsonian Center for Astrophysics, 60 
Garden Street, Cambridge, MA 02138, USA}\\
\altaffiltext{3}{Steward Observatory, University of Arizona, 933 
N Cherry Ave, Tucson, Az, 85721}
\vspace{-0.5cm}
}

\date{Submitted to MNRAS, November 1, 2011\vspace{-0.6cm}}
\begin{document}
\maketitle
\label{firstpage}

\begin{abstract}

It is well-established observationally that the characteristic angular
momentum axis on small scales around AGN, traced by radio jets and the
putative torus, is not well-correlated with the large-scale angular
momentum axis of the host galaxy.  In this paper, we show that such
misalignments arise naturally in high-resolution simulations in which
we follow angular momentum transport and inflows from galaxy to sub-pc
scales near AGN, triggered either during galaxy mergers or by
instabilities in isolated disks.  Sudden misalignments can sometimes
be caused by single massive clumps falling into the center slightly
off-axis, but more generally, they arise even when the gas inflows are
smooth and trace only global gravitational instabilities.  When
several nested, self-gravitating modes are present, the inner ones can
precess and tumble in the potential of the outer modes.  Resonant
angular momentum exchange can flip or re-align the spin of an inner
mode on a short timescale, even without the presence of massive
clumps.  We therefore do not expect that AGN and their host galaxies
will be preferentially aligned, nor should the relative alignment be
an indicator of the AGN fueling mechanism. We discuss implications of
this conclusion for AGN feedback and BH spin evolution.  The
misalignments may mean that even BHs accreting from smooth large-scale disks 
will not be spun up to maximal rotation, and so have more modest radiative efficiencies and inefficient jet formation. 
Even more random orientations/lower spins are possible if there is further, un-resolved clumpiness in the gas, 
and more ordered accretion may occur if the inflow is slower and not self-gravitating.

\end{abstract}

\begin{keywords}
galaxies: active --- quasars: general --- 
galaxies: evolution --- cosmology: theory
\vspace{-1.0cm}
\end{keywords}

\vspace{-1.1cm}
\section{Introduction}
\label{sec:intro}

Understanding accretion is critical for inferring the origin of the
supermassive black hole (BH) population
\citep{soltan82,salucci:bhmf,shankar:bhmf,hopkins:old.age}. 
Most of the BH growth in the Universe is obscured by large columns of
gas and dust, so knowing the behavior of gas on scales
$\sim0.1-100\,$pc is a necessary ingredient in a full model of BH
evolution
\citep{antonucci:1982.torus,antonucci:agn.unification.review,lawrence:receding.torus,
risaliti:seyfert.2.nh.distrib,simpson99:thermal.imaging.of.radio.gal,
willott00:optical.qso.frac.vs.l}.  The discovery of tight
correlations between BH mass and host spheroid properties \citep[e.g.\
mass, velocity dispersion, binding energy;][]{KormendyRichstone95,
magorrian,FM00,Gebhardt00,hopkins:bhfp.obs,hopkins:bhfp.theory,aller:mbh.esph,feoli:bhfp.1}
implies that BH growth is coupled to galaxy formation.  Models widely
invoke some form of feedback from AGN to explain the origin of the
BH-host relations, rapid quenching of star formation in bulges, the
color-magnitude relation, and the cooling flow problem
\citep[e.g.][and references therein]{silkrees:msigma,king:msigma.superfb.1,
king:msigma.superfb.2,dimatteo:msigma,springel:red.galaxies,
hopkins:groups.qso,hopkins:twostage.feedback,croton:sam}.

However, despite these important links, the detailed processes in BH
fueling remain poorly understood.  One critical long-standing puzzle
is the consistent observational finding that there is little or no
correlation between the angular momentum axis of material accreting
onto the BH, and the axis of the host galaxy.  This has been seen with
a number of different tracers, e.g.\ radio jets (expected to align
with the axis of the BH spin or inner accretion disk, but see also 
\citealt{natarajan:1998.jet.angle.vs.bh.spin}) or obscuring AGN
``torii'' defining the plane along which material flows into the inner
accretion disk
\citep[see e.g.][]{keel:1980.seyfert.vs.galaxy.inclination,lawrence:1982.torus.alignment,
ulvestad:1984.radiojet.seyfert.misalignment,
schmitt:1997.radio.alignment.w.host,simcoe:1997.agn.host.alignment,kinney:2000.bh.jet.directions,
gallimore:2006.agn.outflow.gal.alignment,zhang:2009.agn.vs.hubble.type}.\footnote{We stress that this is not necessarily the same as a lack of correlation between obscuration and host galaxy alignment, since 
significant obscuring columns can come from large scales in e.g.\ starbursts or edge-on disks 
\citep{hopkins:qso.all,hopkins:seyferts,
hayward:2011.smg.merger.rt,zakamska:qso.hosts,rigby:qso.hosts,lagos:2011.agn.gal.orientation}.} 
The nuclear disk is misaligned with the larger-scale disk/galaxy inflows; but the latter 
must ultimately be the origin of the former, so this is not trivially expected. 

This misalignment has a number of consequences. It constrains any
model of AGN fueling and has important implications for AGN
obscuration.  Not only does it constrain the origin of the ``torus,''
but misalignments between the inner and outer disk can potentially
result in large covering factors of obscuration \citep[even if the
disks are thin; see
e.g.][]{sanders:1989.warped.qso.disk.ir.emission,nayakshin:2005.warped.disk.obsc,
fruscione:abs.by.warped.disk,hopkins:torus}.  It is critical
for understanding BH spin -- if gas accreted from large scales in
the galaxy conserves its axis of angular momentum as it falls onto the
BH, then almost any high accretion rate event will spin the BH up to
near-maximum ($a\approx0.998$) and align it with the parent
disk/inflow \citep[e.g.][]{volonteri:2005.bh.spin.pred,volonteri:2005.bh.spin.sam,berti:2008.bh.spin.sam}. 
However, if the angular momentum can be randomized on sufficiently small 
mass/timescales (``chaotic accretion''), then not only will the lack of correlation with the host galaxy appear 
\citep{king:2007.align.radio.torus.notgal}, but the typical spins are held low even in 
large accretion events \citep{moderski:1996.chaotic.acc.lowspin,king:2006.chaotic.acc.lowspins}. 
Spin has important subsequent implications for BH-BH mergers and gravitational
wave BH recoil (whether or not BHs will be expelled from the galaxy or
rapidly damp any small recoil motion). And it is believed to be
critical for the production of radio jets
(at least in some scenarios; see \citealt{blandfordznajek:jet.spin,begelman:1984.radio.jet.review}; but also 
compare \citealt{livio:1999.non.spin.jet.from.accdisk}). 
Jets and other AGN feedback sources are of critical importance for quenching
cooling in massive galaxies, shaping the galaxy mass 
function, structuring galaxy clusters, and resolving the ``cooling flow problem.'' 

Unfortunately, it is not generally possible to simultaneously model 
inflows from galactic scales and their behavior on the small scales near the BH that 
are relevant for this problem. 
Analytic models \citep{kawakatu:disk.bhar.model,
nayakshin:forced.stochastic.accretion.model,elitzur:torus.wind} 
are limited by symmetry assumptions 
as well as the fact that these systems are highly non-linear, often chaotic, and 
not in steady-state (with inflow, outflow, star formation, and feedback competing). 
Simulations of galaxies used to follow inflows 
are typically limited to a resolution of several $100\,$pc, much larger than the scales of 
interest \citep{cattaneo:2005.mgr.agn.obsc,
hopkins:lifetimes.methods,hopkins:lifetimes.obscuration}. 
Other simulations which begin on small scales (taking some fixed initial conditions 
for the gas inside of $\lesssim10\,$pc), cannot relate this to the 
larger-scale material from which it must have originated \citep{schartmann:2009.stellar.fb.effects.on.torus,
wada:starburst.torus.model,wada:torus.mol.gas.hydro.sims}. 
Some exciting results have emerged from ``zoom-in'' refinement techniques
\citep[see][]{escala:nuclear.gas.transport.to.msigma,colpi:2007.binary.in.mgrs,
levine2008:nuclear.zoom,mayer:bh.binary.sph.zoom.sim,
dotti:bh.binary.inspiral}, but
computational expense has generally required restrictive assumptions
(e.g.\ turning off cooling and star formation on small scales) or
limited these to single example galaxies at a single instant in time
(preventing statistical statements).

Recently, \citet{hopkins:zoom.sims} attempted to build on these 
experiments to model the angular momentum
transport required for massive BH growth, and carried out a series of
numerical simulations of inflow from galactic to BH
scales.
By ``re-simulating'' the central regions of
galaxies in a series of stages, gas flows can be modeled over a range of galactic scales from
$\sim100\,$kpc to $<0.1\,$pc.  In \citet{hopkins:zoom.sims} we show that
quasar-level inflows ($\sim10\,\msun\,{\rm yr^{-1}}$) arise from global
perturbations such as galaxy mergers and/or secular instabilities,
which (when sufficiently strong) generate a cascade of subsequent
instabilities (of varied morphology), and typically manifest near the
radius of influence of the BH as a thick (torus-like),
lopsided/eccentric gas+stellar disk. In \citet{hopkins:m31.disk} we
discuss evidence for the relics of such disks in nearby galaxies
\citep{lauer93,bender:m31.nuclear.disk.obs, lauer:ngc4486b}. In
\citet{hopkins:inflow.analytics} we discuss the detailed dynamics of
these instabilities and how they drive large inflow rates, and in
\citet{hopkins:torus} discuss their role in the obscuration of AGN.

\begin{figure}
    \centering
    \plotone{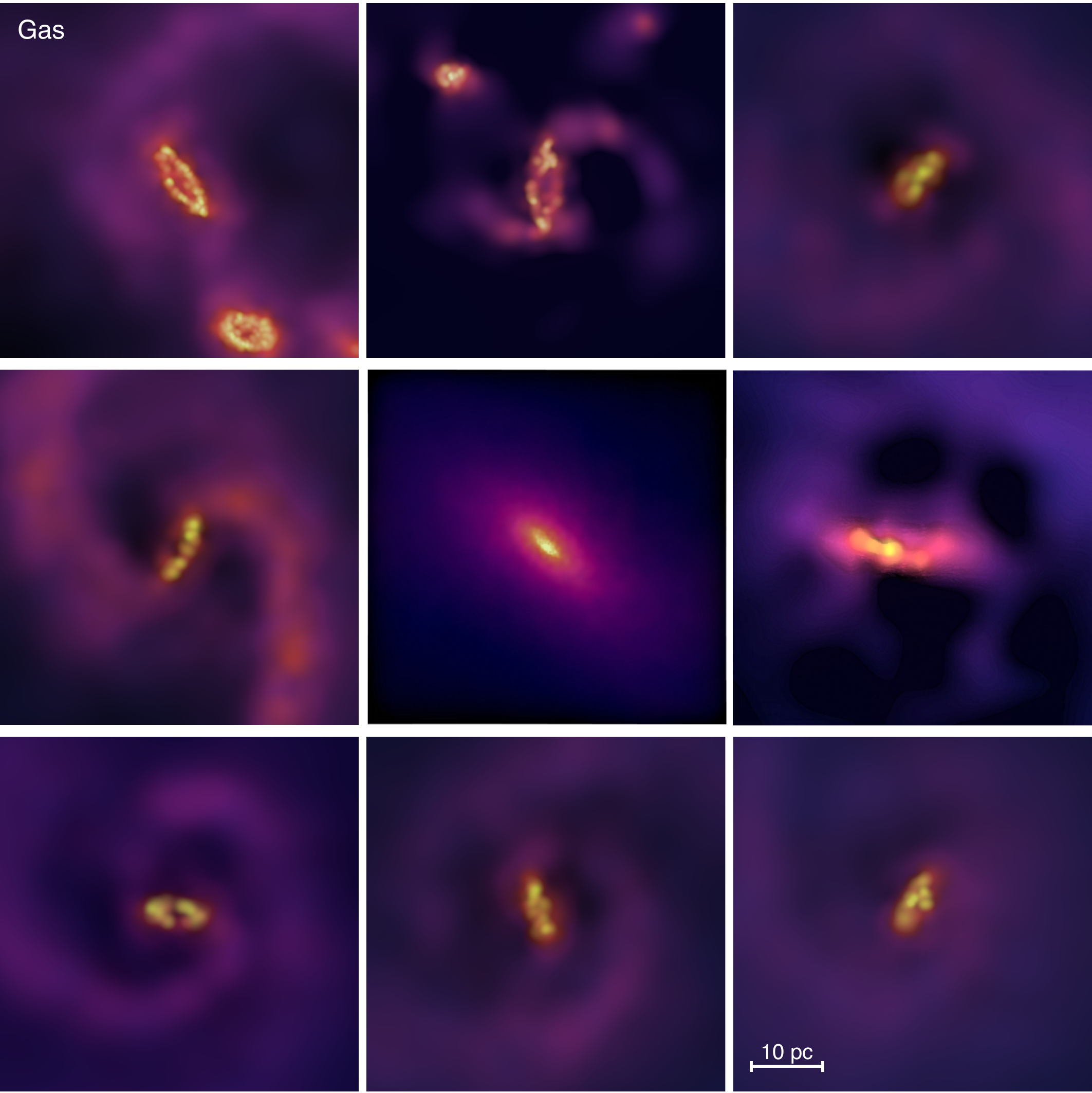}
    \caption{Illustration of twists, warps, and misalignments in 
    some of our simulations (each panel is a different simulation). 
    Projected gas density (intensity) and 
    specific star formation rate (color, increasing from blue through yellow) are 
    shown. Times are chosen randomly near the peak of the BH accretion.
    All are projected ``face on'' to the disk on large 
    scales (angular momentum averaged over the entire box). 
    These are ultra high-resolution ``intermediate-scale'' simulations which can 
    resolve twists and misalignments between 
    the ``torus'' and larger-scale disk. There is frequently such a 
    misalignment or warp between the inflow from larger scale bars-within-bars, 
    and the nuclear disk (e.g.\ middle-left or bottom-left cases), 
    or misalignment driven by the inflow of large clumps from 
    fragmentation of large-scale modes (e.g.\ top-left or top-center cases). 
    \label{fig:torus.zoom.alignment}}
\end{figure}

In this {\em Letter}, we show that the instabilities which
drives inflows in these simulations naturally lead to large misalignments of the nuclear disk with respect to the disk of the host galaxy.  We discuss the implications for BH spin, even in ``maximally conservative'' scenarios where there is no unresolved sub-grid clumpiness in the ISM in our simulations (there almost certainly is such).

\vspace{-0.7cm}
\section{The Simulations}
\label{sec:sims}

The simulations used here are taken from a suite used to study the physics of gas inflow from 
galactic to small scales in \citet{hopkins:zoom.sims,hopkins:inflow.analytics,hopkins:m31.disk,
hopkins:slow.modes,hopkins:cusp.slopes}. The numerical properties of each simulation 
are specifically given in \citet{hopkins:zoom.sims} (Tables~1-3), but we briefly describe them here. 
In order to probe the very large range in spatial and mass scales, we
carry out a series of ``re-simulations.'' First, we simulate the
dynamics on galaxy scales.  Specifically, we use representative
examples of gas-rich galaxy-galaxy merger simulations and isolated,
moderately bar-unstable disk simulations. These are well-resolved down
to $\sim100-500\,$pc.  We use the conditions at these radii (at
several times) as the initial conditions for intermediate-scale
re-simulations of the sub-kpc dynamics. In these re-simulations,
the smaller volume is simulated at higher resolution, allowing us to
resolve the subsequent dynamics down to $\sim10\,$pc scales -- these
re-simulations approximate the nearly instantaneous behavior of the
gas on sub-kpc scales in response to the conditions at $\sim$kpc set
by galaxy-scale dynamics. We then repeat our re-simulation method
to follow the dynamics down to sub-pc scales where the gas begins to
form a standard accretion disk.

Our re-simulations are {\em not} intended to provide an exact realization of the small-scale dynamics of the larger-scale simulation that motivated the initial conditions of each re-simulation (in the manner of particle-splitting or adaptive-mesh refinement techniques). 
Rather, our goal is to identify the dominant mechanism(s) of angular exchange and transport in galactic nuclei and what parameters they depend on.  This approach clearly has limitations, especially at the outer boundaries of the simulations; however, it also has a major advantage. By  not requiring the conditions at small radii to be uniquely set by a larger-scale ``parent'' simulation, we can run a series of simulations with otherwise identical conditions (on that scale) but systematically vary one parameter (e.g., gas fraction or ISM model) over a large dynamic range.  This allows us to identify the physics and galaxy properties that have the biggest effect on gas inflow in galactic nuclei. The diversity of behaviors seen in the simulations, and desire to marginalize over the uncertain ISM physics, makes such a parameter survey critical.

The simulations were performed with the TreeSPH code {\small
GADGET-3} \citep{springel:gadget}; they include stellar disks, bulges, dark 
matter halos, gas and BHs. For this study, we wish to isolate the physics of gas 
inflow and so do not include explicit models for BH feedback (see \S~\ref{sec:discussion}). 
The simulations include gas cooling and star formation, with gas
forming stars at a rate $\dot \rho_{\ast} \propto \rho^{3/2}$ 
motivated by observations \citep{kennicutt98}, and normalized so 
a MW-like galaxy has total $\dot{M}_{\ast}\approx1\,M_{\sun} \, {\rm yr^{-1}}$.
Varying the exact slope or normalization of this assumption has no qualitative
affect on our conclusions.  
Because we cannot resolve the detailed processes of supernovae
explosions, stellar winds, and radiative feedback, feedback from stars
is modeled with an effective equation of state 
\citep{springel:multiphase}. In this model, feedback is assumed to
generate a non-thermal (turbulent, in reality) sound speed that
depends on the local star formation rate, and thus the gas density; 
the results shown span a wide range in this ``effective sound speed'' 
without any strong dependence on the exact value
(detailed comparisons of the effects on morphology and inflow are shown in 
\citealt{hopkins:zoom.sims}; comparisons of mode growth and torques in \citealt{hopkins:inflow.analytics}). 
More detailed comparison with the explicit stellar feedback models presented in 
\citet{hopkins:rad.pressure.sf.fb,hopkins:fb.ism.prop,hopkins:stellar.fb.winds} 
will be the subject of future work.

We ``begin'' with galaxy-scale simulations that motivate the initial conditions chosen
for the smaller-scale re-simulation calculations. These include galaxy-galaxy mergers, and
isolated bar-(un)stable disks. These simulations have
$0.5\times10^{6}$ particles and spatial resolution of
$50\,$pc \citep[details in][]{dimatteo:msigma,robertson:msigma.evolution,
  cox:kinematics,younger:minor.mergers,hopkins:disk.survival}; a subset have 
  $\sim10^{7}$ particles and $20\,$pc resolution.
From this suite we select representative simulations
of gas-rich major mergers of Milky-Way mass galaxies 
and their isolated bar-unstable analogues, to
provide the basis for our re-simulations. 
Small variations in the orbits or the structural properties of the galaxies
will change the details of the tidal and bar features on large scales; however, 
we show in \citet{hopkins:zoom.sims} that the precise details of these large-scale
simulations do not {\em instantaneously} alter the dynamics on small scales 
(see Figs.~A2 \&\ A3 therein). 
Rather, the local dynamics depends on global parameters such as the gas mass channeled
into the central region, relative to the pre-existing bulge, disk, and
black hole mass (set, of course, by the large-scale inflows, but once set, robust to variations in the 
details of that inflow structure). 

Following gas down to the BH accretion disk requires much higher
spatial resolution than is achievable in the galaxy-scale simulations. We 
therefore select snapshots from the galaxy-scale simulations at key epochs and 
isolate the central $0.1-1$\,kpc region which contains
most of the gas driven in from large scales
(typical $\sim10^{10}\,\msun$ in gas, over scale-length $\sim0.3-0.5\,$kpc).
From this mass distribution, we then re-populate the gas in the
central regions at much higher resolution, and simulate the dynamics
for several local dynamical times.  These ``intermediate-scale'' simulations involve $10^{6}$
particles, with a resolution of a few pc and particle masses of
$\approx 10^{4}\,\msun$.  We have run $\sim50$ such re-simulations,
corresponding to variations in the global system properties, the model
of star formation and feedback, and the exact time in the larger-scale
dynamics at which the re-simulation occurs.
\citet{hopkins:zoom.sims} present tests of this
re-simulation approach and show that it is reasonably robust for this
problem.  This is largely because, for gas-rich disky systems, the
central $\sim 300$ pc becomes strongly self-gravitating, generating
instabilities that dominate the subsequent dynamics.

We repeat our re-simulation process once more, using
the central $\sim10-30\,$pc of the first re-simulations to initialize
a new set of ``small-scale'' simulations.  These typically have
$\sim10^{6}-10^{7}$ particles,
a spatial resolution of $0.1\,$pc, and a particle mass
$\approx100\,\msun$. We carried out $\sim50$ such simulations to test
the robustness of our conclusions and survey the parameter space of
galaxy properties.  These final re-simulations are evolved for
$\sim10^{7}$ years -- many dynamical times at $0.1$\,pc, but very
short relative to the dynamical times of the larger-scale parent
simulations. 

To check that our re-simulation approach has not introduced any artificial behavior, 
we have run a small number of higher resolution ``bridging'' simulations. These 
result in slightly worse ultimate spatial resolution than the net effect of the ``re-simulations,'' but they obviate the need for 
the re-simulation and bridge the scales of the above simulation suites. These include 6 simulations 
on galaxy scales (3 mergers, 3 isolated disks) with $>10^{7}$ gas particles and $10\,$pc softening lengths. 
While not quite as high-resolution as our ``intermediate-scale'' re-simulation runs, 
these provide an important check on the results of the latter and are run self-consistently for $4\times10^{9}\,$yr.   
We have followed the same procedure on small scales: running 5 ``intermediate-scale'' simulations (with a 
range of gas fraction and bulge-to-disk ratio) with $>10^{7}$ gas particles and softening of $\sim0.3\,$pc; 
these extend from scales $\sim 0.3-1000$ pc and are run for $2\times10^{8}\,$yr.
In \citet{hopkins:zoom.sims} and \citet{hopkins:inflow.analytics} we explicitly compare the results of these 
simulations with those of our ``re-simulations'' at the dynamic range where they overlap, and find they are very similar 
(see e.g.\ the discussion and Figs.~9-13 \&\ A4 in \citealt{hopkins:zoom.sims} 
and Fig.~8 in \citealt{hopkins:inflow.analytics}), supporting the methodology used for most of our calculations.

We note that recent studies comparing cosmological simulations done
with  {\small GADGET} and the new moving mesh code {\small AREPO}
\citep{springel:arepo} have called into question the reliability of
smoothed particle hydrodynamics (SPH) for some problems related to
galaxy formation in a cosmological context \citep{vogelsberger:2011.arepo.vs.gadget.cosmo,
sijacki:2011.gadget.arepo.hydro.tests,keres:2011.arepo.gadget.disk.angmom,bauer:2011.sph.vs.arepo.shocks}.  However, we have also
performed idealized simulations of mergers between individual
galaxies and found excellent agreement between {\small GADGET}
and {\small AREPO} for e.g. gas-inflow rates, star formation
histories, and the mass in the ensuing starbursts
\citep{hayward:arepo.gadget.mergers}.  Simulations of this type circumvent many
of the issues with SPH by characterizing the gas on small
scales with an effective equation of state (as in the present study), rather than
attempting to resolve the various gas phases explicitly. The discrepancies 
above are also minimized when the flows of interest are supersonic (as opposed to 
sub-sonic), which is very much the case here \citep{kitsionas:2009.grid.sph.compare.turbulence,price:2010.grid.sph.compare.turbulence,bauer:2011.sph.vs.arepo.shocks}. 
We have also performed direct resolution studies of simulations at each ``scale'' 
(with up to $168$ times as many particles) and find good convergence 
(see e.g.\ \S~A1 and Fig.~A1 in \citealt{hopkins:zoom.sims} and 
Fig.~4 in \citealt{hopkins:inflow.analytics}).

\begin{figure}
    \centering
    \plotone{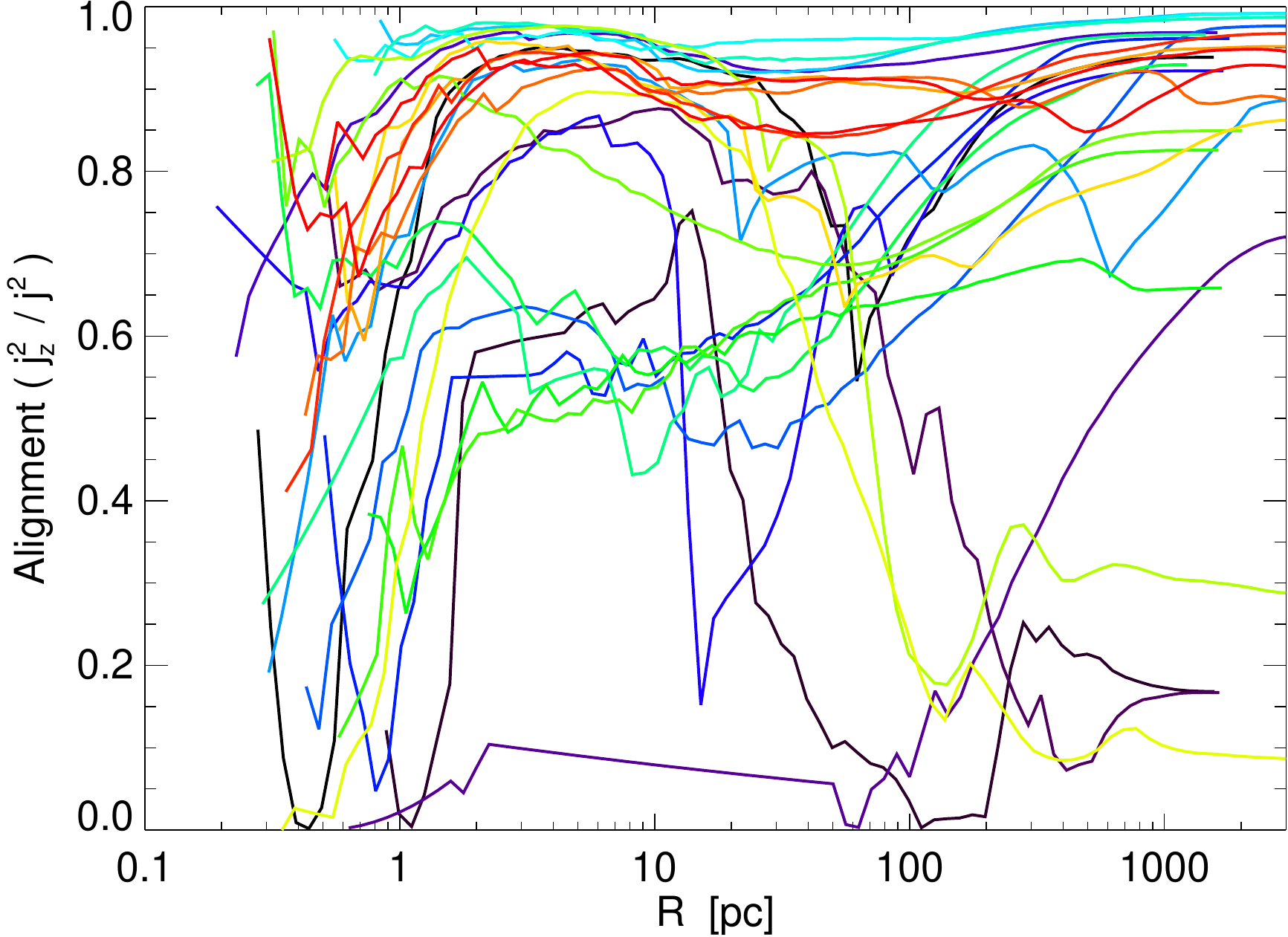}
    \caption{Alignment of the gaseous disks as a function of radii, 
    across our sample of simulations (each line is one simulation 
    chosen near the peak in inflow). 
    We quantify (mis)alignment as the ratio of $j_{z}^{2}/j^{2}$, 
    where $j$ is the total angular momentum vector of the gas within 
    an annulus around radius $R$, and the $z$ axis is (by definition) 
    the angular momentum axis of the entire galaxy. 
    Within $<10\,$pc, there is relatively weak correlation between 
    the inner and outer disk angles. 
    \label{fig:alignment}}
\end{figure}

\vspace{-0.9cm}
\section{Results}
\label{sec:results}

Figure~\ref{fig:torus.zoom.alignment} shows the central tens of pc in several of our ``intermediate-scale'' bridging simulations, in which inflows are followed from $\sim1-1000\,$pc scales. Of course, resolving those larger scales and the resulting inflow means that the resolution on these scales is not quite as good as our ``small-scale'' runs, but the $\sim10\,$pc scale disk is marginally resolved (in length/mass; the vertical/internal structure is not resolved below these sizes in these runs).\footnote{See Fig.~10 in \citealt{hopkins:zoom.sims}, which shows the vertical scale heights as a function of radius in these and the others of our simulation suite, compared to our SPH smoothing. At $\sim10\,$pc, the disks have scale heights from $\sim 1-3\,$pc, compared to a softening of $\approx0.3$\,pc, so the internal structure cannot be resolved at smaller radii. The true ``nuclear scale'' re-simulations have resolution of $\approx0.1\,$pc, and so resolve $h/R$ to $\sim1-3\,$pc.} There are clear cases where the inflows from sub-kpc scale bars map onto the disk at the BH radius of influence, but with a very significant misalignment between the two. 

Figures~\ref{fig:alignment} \&\ \ref{fig:angle.summary} quantify the degree of misalignment of the nuclear regions in the simulations. Since the observable quantity is generally the absolute value of the misalignment, we plot $j_{z}^{2}/j^{2}$ (i.e.\ $\cos^{2}{\theta}$), where $j$ is the specific angular momentum in an annulus, and the $z$ axis is the axis of the net angular momentum vector of the entire galaxy. Figure~\ref{fig:alignment} plots this as a function of radius, at a given time in each simulation. Note that there are misalignments at all radii from $\sim0.1$\,pc to $\sim10$\,kpc (although the cases where there are significant misalignments on $>$kpc scales are generally galaxy mergers). Figure~\ref{fig:angle.summary} plots the cumulative distribution of this quantity at a fixed small radius, summed over all times and the entire ensemble of simulations. The misalignments on small scales are somewhere between pure random and pure alignment.


\begin{figure}
    \centering
    \plotone{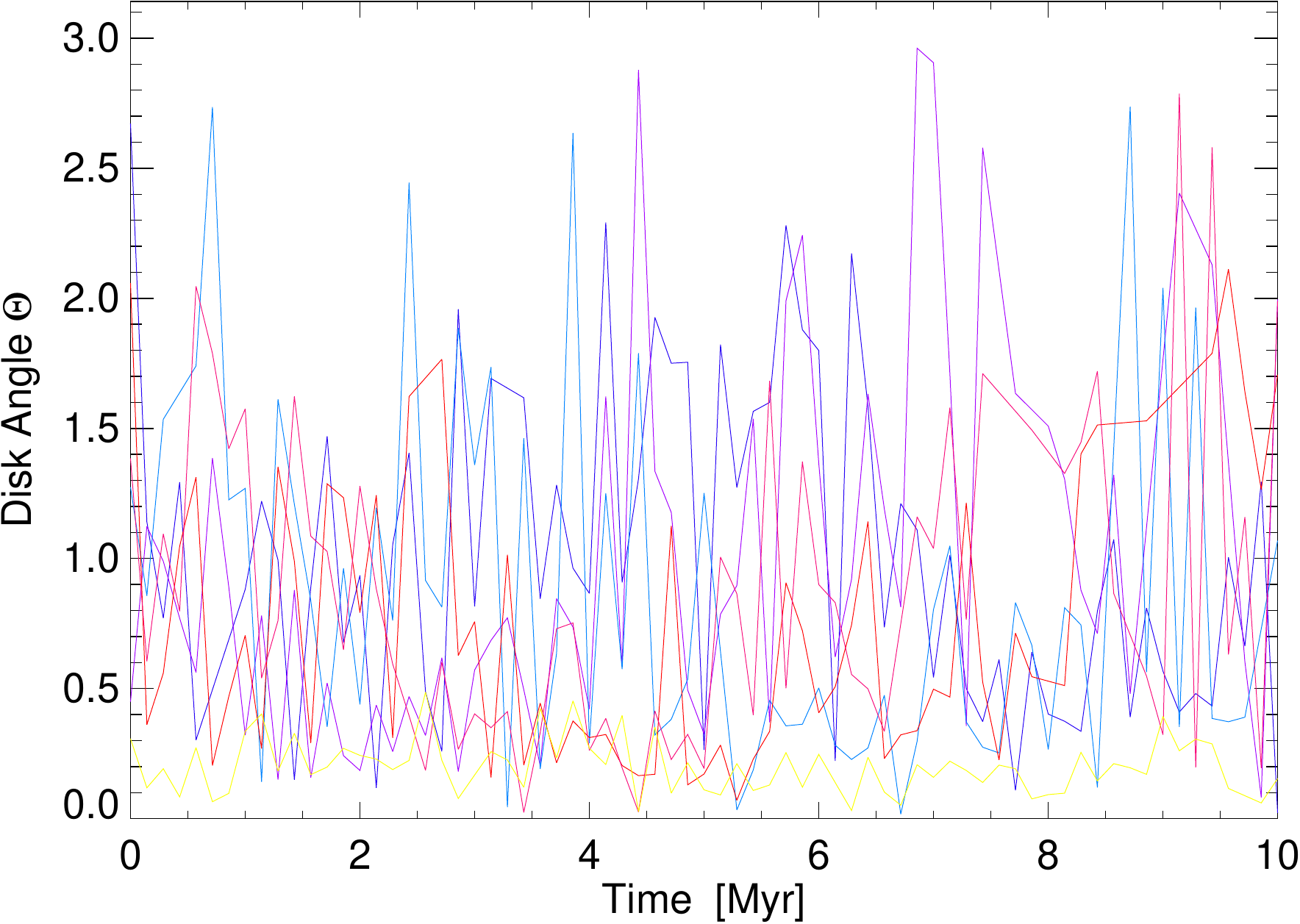}
    \caption{Polar angle $\Theta$ between the nuclear disk at our smallest resolved 
    scale and the initial (uniform) angular momentum axis of the entire system, 
    as a function of time (same simulations as Figure~\ref{fig:alignment}; for clarity 
    and to show short time-scale variability, we plot only a small fraction of the simulated time). 
    The BHs here are accreting at $\sim10\%$ of Eddington; at this rate, systems whose inflow angular 
    momentum axis is effectively random over $\sim10^{7}\,$yr timescales (the duration shown here) will be spun down to low/modest 
    spin values; this includes most of our simulations. If they accrete at Eddington, the relevant timescale is $\sim$Myr; this includes only the most rapidly variable simulations.
    \label{fig:spin}}
\end{figure}

Figure~\ref{fig:spin} illustrates the time evolution of the inflow axis. We plot the evolution of the central angular momentum orientation as a function of time: specifically the angle $\theta$ defined between ${\bf j(t)}$ (total angular momentum vector within $5$ smoothing lengths of the BH -- a couple pc) and the (fixed) $z$-axis (initial ${\bf j}=j\,\hat{z}$). Variation in $\phi$ (azimuthal angle) is much more rapid, but is less significant physically (since systems are axisymmetric to lowest order $\phi$ variation reflects lopsided/eccentric modes). There is large time-variability. The most extreme cases exhibit several ``flips'' with $\theta>\pi/2$ (anti-alignment of the disk with its original inclination). 

Fully understanding how this affects BH spin would require a number of sub-grid assumptions beyond the model here 
\citep[see e.g.][]{fanidakis:2011.bh.spin.sam}. Even if we assume that the gas retains its angular momentum axis below the resolution limit, we need to follow the orientation and magnitude of the BH spin as a function of time, which evolves as, at first, Lens-Thirring alignment forces the inner accretion disk to either align or anti-align (depending on whether ${\bf j}_{\rm disk}\cdot {\bf j}_{\rm BH}>0$ or $<0$, respectively) inside of some warp radius $R_{\rm warp}$, and then the torques associated with this eventually re-orient the BH spin in alignment with the disk \citep{bardeen:1975.lens.thirring.alignment}. For a given ``event,'' full alignment will occur if $\cos{\theta}>-J_{d}/2\,J_{\rm BH}$ (where $\theta$ is the original angle between the BH and disk, $J_{\rm BH}$ the BH spin angular momentum, and $J_{d}$ is approximately the interior disk angular momentum passing through the warp region; \citealt{scheuer:1996.bh.acc.disk.alignment,king:2005.bh.acc.alignment.criterion}). But this also depends on the sub-structure, dynamics, and properties of the internal $\alpha$-disk, well below our best-case resolution \citep[see][]{kumar:1985.misalignment.bh.spin.viscous.disk}. However, crudely speaking, for typical $\alpha$-disk models, this translates to a criterion on the mass accreted in a given ``event'' with coherent angular momentum: if the angular momentum remains coherent over a timescale long enough for the BH to accrete some fraction \citep[typically a few percent;][]{lodato:2006.misaligned.acc.bh.spin,perego:2009.spin.evol.accretion} of its mass, then the spin will re-orient to align (even if initially retrograde) and most of the accretion will go to spinning up the BH. If the inflow angular momentum is incoherent on this time/mass scale, however, the spin undergoes a random walk with decreasing magnitude \citep{king:2006.chaotic.acc.lowspins,king:2008.fragmentation.chaotic.accretion}. If the BH is accreting at a fraction $\lambda$ times Eddington, this corresponds to a physical timescale of $\sim\lambda^{-1}\,10^{6}$\,yr. Consider this in Fig.~\ref{fig:spin}. If the accretion is sufficiently rapid ($\lambda\approx1$), then only the most extreme simulated variability will be sufficient to give very low spins. However, for the more typical $\lambda\approx0.1$ (coherence time $10^{7}\,$yr) in observed systems \citep{kollmeier:eddington.ratios,hickox:multiwavelength.agn,hopkins:mdot.dist,
trump:2009.type1.agn.mdot.limits} and actually calculated (via the inflow rate into $<0.1$\,pc) in these simulations, a large fraction of the simulations have sufficient resolved precession in their inflow angular momenta to produce a ``random walk'' spin behavior.

\begin{figure}
    \centering
    \plotone{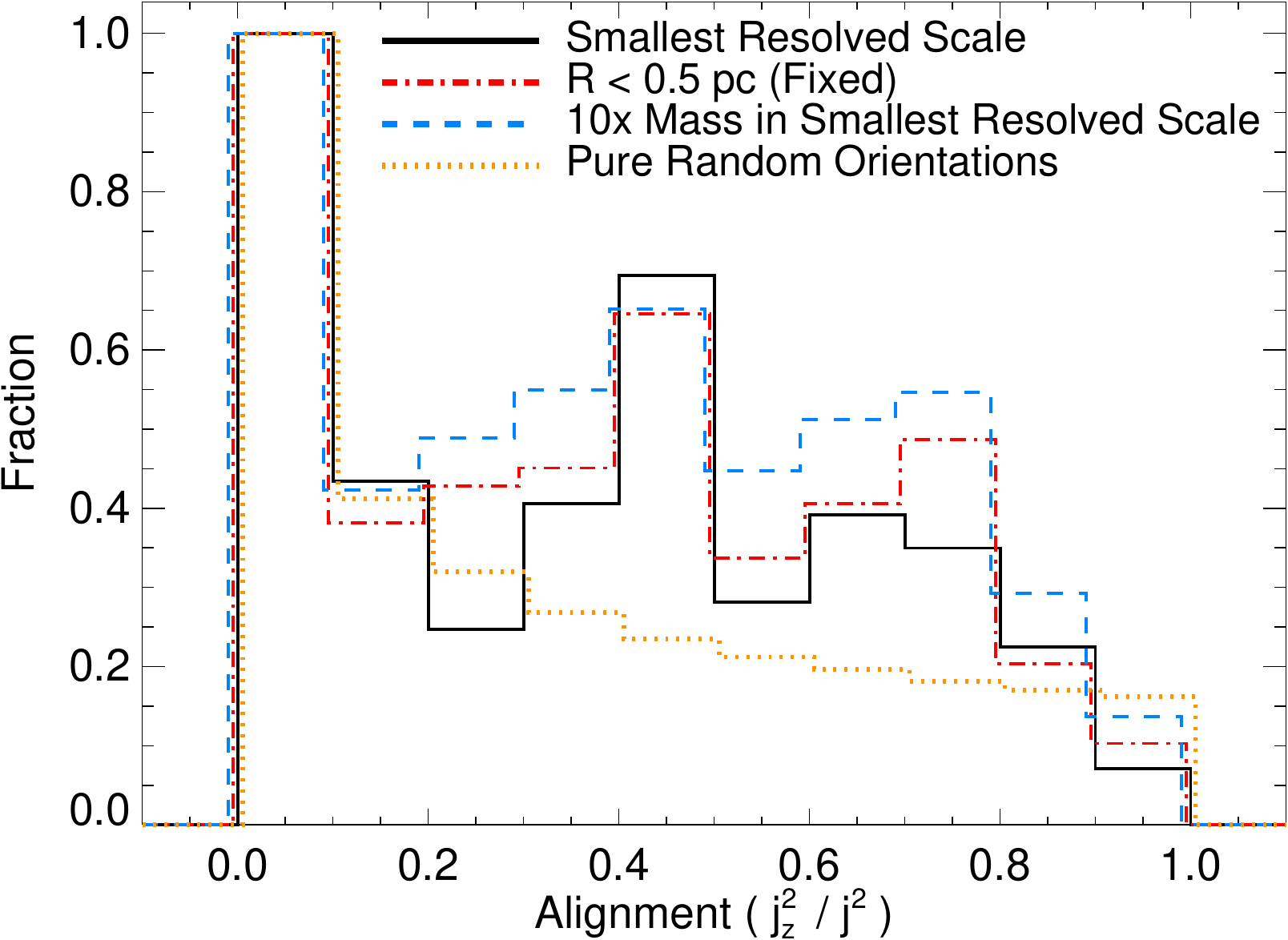}
    \caption{Distribution (averaged over time, and over the ensemble of simulations) of the alignment/angle ($j_{z}^{2}/j^{2} = \cos^{2}{\theta}$) between the nuclear disk and angular momentum axis of the large-scale system (where $j_{z}=j$, by definition). We measure this in three radii: the smallest resolved radii used in Figure~\ref{fig:spin}, a fixed physical radius of $0.5\,$pc, and a radius enclosing ten times the gas mass of that enclosed in the smallest resolved radius. The three agree reasonably well, suggesting the results are robust at small radii (though this clearly reaches the limits of reliable resolution effects). To compare, we show the distribution which would be obtained if the orientations were purely random and uniformly distributed over the sky. Pure alignment would be a $\delta$-function at $j_{z}^{2}/j^{2}=1$. The distribution is closer to random than to pure alignment; still, there is an obvious tendency for more alignment than in the pure random case. There may also be a weak excess of misalignments near $\cos^{2}{\theta}\sim 0.4-0.5$, but this is  marginally significant.
    \label{fig:angle.summary}}
\end{figure}

\vspace{-0.7cm}
\section{Discussion}
\label{sec:discussion}

Using high-resolution simulations of gas inflows from galaxy to sub-pc scales around AGN, 
we study the evolution of BH-host galaxy alignments. 
We predict only a weak correlation between the nuclear axis and the large-scale 
disk axis. If anything, this is a lower limit to the typical degree of ``randomness'' in alignment, as more clumpy 
star formation or infall from recycled stellar wind material can increase the variation in 
orientations. Twists and misalignments, therefore, can explain the random alignment of AGN disks 
relative to their host galaxies. A warped or twisted disk may also yield large covering angles towards 
the BH even when the disk 
itself is thin, although we argue in \citet{hopkins:torus} that this is not alone sufficient to explain observed obscuration (the ``torus'' must also be geometrically thick). 

These misalignments occur for at least two reasons.
First, there are cases where large-scale 
fragmentation occurs in the gas (part of a spiral arm or other instability fragments and sinks to the 
center), which can dramatically change the nuclear gas angular momentum 
content \citep[see also][]{nayakshin:forced.stochastic.accretion.model,levine:sim.mdot.pwrspectrum,
king:2008.fragmentation.chaotic.accretion}. 
And second, even in perfectly smooth flows, 
it is well-known that secondary bars in the presence of dissipative processes 
(i.e.\ gas) will tend to de-couple their angular momentum from the primary 
bar \citep[e.g.][and 
references therein]{heller:secondary.bar.instability}. 
Inflow and dissipation lead to runaway strengthening of the 
inner mode, which populates various chaotic orbit families and exchanges angular 
momentum with the outer mode, de-coupling the inner mode angular momentum 
and orbit plane from that of the outer mode \citep{hasan:1990.bar.cmc.chaos,heller:1996.dynamics.nuclear.rings,maciejewski:2000.bar.bar.orbits}. The inner mode 
precesses or tumbles in three dimensions relative to the outer 
mode frame, a phenomenon seen in a large number of simulations 
\citep{shlosman:nested.bar.evol,elzant:nested.bar.scales,
maciejewski:nested.bar.models,englmaier:nested.bar.decoupling}
and observed double (and even triple) bars \citep{shaw:1995.nuclear.bar.inflow,friedli:1993.bars.win.bars.bulges,friedli:1996.disk.obs.multiple.substruct,erwin:1999.triple.bar.structure.obs,
laine:nested.bars.in.seyferts,erwin:double.bar.obs}. 
These processes are common in our simulations, especially in the complicated triaxial potential of realistic merger-formed bulges. 

An analogous process also occurs 
here with the inner lopsided disk, at the inner radius (ILR) of the outer bar 
(itself, in several cases, the ``inner'' of a double bar). \citet{hopkins:inflow.analytics} show in both these simulations and analytic calculations that angular momentum exchange in the gas in the central regions (inside the BH radius of influence but outside the viscous accretion disk) can be strongly dominated by super-sonic gas shocks surrounding strong torquing regions in the stellar nuclear disk with  lopsided/eccentric ($m=1$) modes \citep[see also][]{salow:nuclear.disk.models,jacobs:longlived.lopsided.disk.modes,sambhus:m31.nuclear.disk.model,bacon:m31.disk}. These modes can resonantly exchange angular momentum with the pattern at larger radii in the manner of nested bars, leading (in plane) to possible reversals and counterrotation of the pattern, which in turn reverses the sense of torques on the gas. If the mode is strong enough, the exchange in strong shock regions can be large enough to change the gas angular momentum by an order-unity factor. Generally, as the gas approaches the mode, it experiences a sudden, strong resonant torque, shortly followed by or accompanying a strong shock that dissipates its energy. When the torques are sufficiently strong the gas falls in on a nearly radial orbit along the pattern; the small ``residual'' angular momentum can have different signs depending on the instantaneous pattern speed, precession rate, and resonance structure of the mode (all of which continuously evolve). Moreover, we show in \citet{hopkins:torus} that when a sufficiently strong $m=1$ mode appears, the inner disk becomes vulnerable to the ``firehose instability,'' and self-excites large vertical bending modes. The linear derivation of those modes therein suggests that they have both large growth rates and order-unity saturation amplitudes (i.e.\ drive order-unity fluctuations in $\theta$ in Fig.~\ref{fig:spin}); if there is a significant population of stars in the nucleus on retrograde orbits (from, say, previous accretion episodes), the growth rate and saturation amplitude of these modes is greatly enhanced \citep{sellwood:1994.counterrot.bend.instab,davies:1997.counterrot.disk.instab}. 
We should note that, although the evolution in Fig.~\ref{fig:spin} is extremely ``rapid'' relative to e.g.\ 
secular processes a $>$kpc radii, it is still indeed secular: at $1\,$pc around a BH of $10^{7}-10^{8}\,\msun$, $1\,$Myr represents $\sim100-1000$ dynamical times, so the relevant resonant effects can collectively operate over a very large number of orbital periods. All of these processes become more prominent as the nuclear gas is 
more strongly self-gravitating; so they may operate progressively more efficiently 
in higher-accretion rate AGN. They do, however, rely on a complicated interaction between collisionless and collisional material; as such, many will not appear in simulations that do not include ``live'' star formation in the disk \citep[compare e.g.][]{escala:nuclear.gas.transport.to.msigma,colpi:2007.binary.in.mgrs}. At least some observed AGN (with e.g.\ jets and maser mapping of their nuclear regions) appear to have such multiple misalignments corresponding to structures (nested bars) in their hosts \citep[see e.g.][]{greenhill:1997.maser.twist}. 

We do not predict perfectly random alignments, as there is still some bias towards similar axes obvious in Figure~\ref{fig:angle.summary} \citep[suggested in observations as well, in][]{battye:2009.radio.alignment.optical,shen:2010.torus.alignment,lagos:2011.agn.gal.orientation}. 
Interestingly, there is also a suggestion in Figures~\ref{fig:alignment} \&\ \ref{fig:angle.summary} of a preference for misalignments of $\cos^{2}(\theta)\sim1/2$ ($\theta = (\pm\pi/4,\,\pm3\pi/4)$). These relative alignments reflect quasi-stable potential surfaces, for example, for an inner gas disk in a tumbling prolate quasi-spherical potential; they also form the backbone of ``X-shaped'' (and some ``peanut-shaped'') bulges formed by bar ``buckling'' after the vertical motions are pumped by resonances in the presence of a nuclear mass concentration or secondary bar, like those seen here \citep{tohline:1982.gas.disk.orientation.in.prolate.gal,pfenniger:bar.dynamics,
pfenniger:1990.bar.buckling.dissipation}. It is not surprising, then, that they form the upper envelope for the misalignments seen in the more ``quiescent'' models here. The much larger mis-alignments seen in Fig.~\ref{fig:alignment} and most dramatic ``flips'' seen in Fig.~\ref{fig:spin}, on the other hand, tend to arise from the action of large clumps/fragmentation. 

This can have important implications for the spin evolution of BHs. If there are no further twists or clumpy structure beyond what is resolved here, the resulting spins will in some cases be maximal, but in a large fraction of our simulations would be modest -- changes in alignment on sufficiently rapid timescale will lead to rapid BH growth when the orbits are prograde, but then suddenly drops when the orbits are retrograde, producing spins in the range $|a|\sim0.1-0.9$ \citep[see e.g.][]{king:2006.chaotic.acc.lowspins}. As these authors and others have noted, intermediate spins are interesting because they imply modest radiative efficiencies \citep[$\epsilon\sim0.05-0.2$, 
which may be favored by BH luminosity density constraints;][]{wang:2009.rad.eff.vs.z}, and reduce by a factor of several the fraction of maximal recoil ``kicks'' with $\Delta v\gtrsim1000\,{\rm km\,s^{-1}}$ in major BH-BH mergers \citep{van-meter:2010.recoil.kick.formula,kesden:2010.relativistic.kick.suppression}, although this depends on the orientation of the orbital angular momentum which may have preferred configurations \citep[e.g.][]{bogdanovic:2007.spin.alignment.shareddisk.merger}, and is itself coupled to the accretion history and feedback efficiency in mergers \citep[e.g.][]{dotti:2010.spin.evol.recoil.fx,blecha:2011.recoil.bh.merger.model}. It has also been suggested that radio jet power may be modest except at near-maximal spins \citep[e.g.][]{tchekhovskoy:bh.spin.vs.radio.pwr}.


If real BHs have very low spins, $|a|\lesssim0.2$, some additional sub-grid processes are required beyond just what we resolve here. Either unresolved sub-grid clumpiness in the ISM that would lead to more ``chaotic'' accretion by increasing the randomness of the disk orientations on small mass scales as individual clumps are accreted \citep{king:2008.fragmentation.chaotic.accretion}, or further twists/bends/misalignments continuing into the $\alpha$ disk \citep[see e.g.][]{kondratko:3079.acc.disk.maser,greenhill:circinus.acc.disk,kinney:2000.bh.jet.directions}. There may also be resonant exchanges associated with the pairing process in BH mergers, some of which may be important to resolve the ``last parsec problem'' \citep[][]{colpi:2007.binary.in.mgrs,dotti:bh.binary.inspiral,nixon:2011.align.binary.acc.disk,nixon:2011.align.binary.fx.bh.mgrs}. On the other hand, if near-maximal spins are the norm, then it suggests that inflow is somewhat less random than what we find here; some other process, such as slower, more extended accretion from low-density diffuse gas which is not gravitationally unstable, may dominate.



Misalignments can also have dramatic implications for BH feedback. An outer disk which is misaligned with the inner disk, especially one which has multiple ``twists,'' presents a larger ``working surface'' on which AGN feedback may couple (as opposed to an AGN in a single, thin disk). Moreover, many feedback mechanisms are predicted to have preferential alignments corresponding to the spin or nuclear gas disk -- radio jets and ionization cones being preferentially polar, broad absorption line winds preferentially planar. If the inner disk precesses rapidly, these mechanisms might appear effectively isotropic to the gas at larger scales in the galaxy.

The results here are reminiscent of those of
\citet{barnes.hernquist.91,barneshernquist96} on somewhat larger scales. They
found that the angular momentum of gas flowing into the nucleus of a
merger remnant can lose its memory of the initial direction of the
disk angular momentum.\footnote{On super-galactic scales, compare e.g.\ \citet{bett:2011.halo.spin.flips}.}
On this basis, they argued that kinematically
decoupled cores in elliptical galaxies may originate in this
manner \citep{hernquist:kinematic.subsystems,cox:kinematics,
hoffman:mgr.orbit.structure.vs.fgas}.  In particular, 
frames from their animated sequences of
their mergers \citep{barnes:1998.mgr.gas.vhstape} 
often display similarities to the images shown
in Figure~\ref{fig:torus.zoom.alignment}.

The results here represent a first study of inflows in a relatively ``smooth'' medium. In future work, we will extend the models here to include the effects of realistic stellar feedback, star formation, and ISM structure, as well as more detailed physical models for AGN feedback. Stellar feedback should always be present in some form, and may (as discussed above) further enhance the ``randomness'' of the angular momentum on small scales. However, the microphysics of star formation and stellar feedback in the vicinity of even a quiescent BH, let alone a rapidly accreting QSO, are quite uncertain. AGN feedback is potentially important during phases of rapid accretion, but this is less clear -- it may be that the effective duty cycle of strong feedback is such that it is not dynamically important for the angular momentum evolution of accreted material during the time when most of the mass is actually accreted (as, when it becomes strong, it suppresses subsequent accretion). This will, of course, depend on the specific feedback mechanisms. We also specifically avoid the BH-BH merger stage, choosing to focus instead on the more simplified case where there is a single BH in the galaxy nucleus. Certainly, ongoing pair merging may introduce additional misalignments (and will have important spin effects, noted above), but since misalignments are observed even in quiescent, isolated systems, their origin must be more general.

\vspace{-0.7cm}
\acknowledgments 
We thank Massimo Dotti and the anonymous referee for comments and suggestions that greatly improved this manuscript, in particular in the discussion of BH spin. We also thank Eliot Quataert and Robert Antonucci for helpful discussions 
in the development of this work. Support for PFH was provided by NASA through Einstein Postdoctoral Fellowship Award Number PF1-120083. DN acknowledges support from the NSF via grant AST-1009452.


\bibliography{/Users/phopkins/Documents/lars_galaxies/papers/ms}

\begin{thebibliography}{150}
\expandafter\ifx\csname natexlab\endcsname\relax\def\natexlab#1{#1}\fi

\bibitem[{{Aller} \& {Richstone}(2007)}]{aller:mbh.esph}
{Aller}, M.~C., \& {Richstone}, D.~O. 2007, \apj, 665, 120

\bibitem[{{Antonucci}(1993)}]{antonucci:agn.unification.review}
{Antonucci}, R. 1993, \araa, 31, 473

\bibitem[{{Antonucci}(1982)}]{antonucci:1982.torus}
{Antonucci}, R.~R.~J. 1982, \nat, 299, 605

\bibitem[{{Bacon} {et~al.}(2001){Bacon}, {Emsellem}, {Combes}, {Copin},
  {Monnet}, \& {Martin}}]{bacon:m31.disk}
{Bacon}, R., {Emsellem}, E., {Combes}, F., {Copin}, Y., {Monnet}, G., \&
  {Martin}, P. 2001, \aap, 371, 409

\bibitem[{{Bardeen} \&
  {Petterson}(1975)}]{bardeen:1975.lens.thirring.alignment}
{Bardeen}, J.~M., \& {Petterson}, J.~A. 1975, \apjl, 195, L65

\bibitem[{{Barnes} \& {Hernquist}(1996)}]{barneshernquist96}
{Barnes}, J.~E., \& {Hernquist}, L. 1996, \apj, 471, 115

\bibitem[{{Barnes} \& {Hernquist}(1998)}]{barnes:1998.mgr.gas.vhstape}
---. 1998, \apj, 495, 187

\bibitem[{{Barnes} \& {Hernquist}(1991)}]{barnes.hernquist.91}
{Barnes}, J.~E., \& {Hernquist}, L.~E. 1991, \apjl, 370, L65

\bibitem[{{Battye} \& {Browne}(2009)}]{battye:2009.radio.alignment.optical}
{Battye}, R.~A., \& {Browne}, I.~W.~A. 2009, \mnras, 399, 1888

\bibitem[{{Bauer} \& {Springel}(2011)}]{bauer:2011.sph.vs.arepo.shocks}
{Bauer}, A., \& {Springel}, V. 2011, \mnras, in press, arXiv:1109.4413

\bibitem[{{Begelman} {et~al.}(1984){Begelman}, {Blandford}, \&
  {Rees}}]{begelman:1984.radio.jet.review}
{Begelman}, M.~C., {Blandford}, R.~D., \& {Rees}, M.~J. 1984, Reviews of Modern
  Physics, 56, 255

\bibitem[{{Bender} {et~al.}(2005)}]{bender:m31.nuclear.disk.obs}
{Bender}, R., {et~al.} 2005, \apj, 631, 280

\bibitem[{{Berti} \& {Volonteri}(2008)}]{berti:2008.bh.spin.sam}
{Berti}, E., \& {Volonteri}, M. 2008, \apj, 684, 822

\bibitem[{{Bett} \& {Frenk}(2011)}]{bett:2011.halo.spin.flips}
{Bett}, P.~E., \& {Frenk}, C.~S. 2011, \mnras, in press, arXiv:1104.0935

\bibitem[{{Blandford} \& {Znajek}(1977)}]{blandfordznajek:jet.spin}
{Blandford}, R.~D., \& {Znajek}, R.~L. 1977, \mnras, 179, 433

\bibitem[{{Blecha} {et~al.}(2011){Blecha}, {Cox}, {Loeb}, \&
  {Hernquist}}]{blecha:2011.recoil.bh.merger.model}
{Blecha}, L., {Cox}, T.~J., {Loeb}, A., \& {Hernquist}, L. 2011, \apj, in
  press, arXiv:1103.3701

\bibitem[{{Bogdanovi{\'c}} {et~al.}(2007){Bogdanovi{\'c}}, {Reynolds}, \&
  {Miller}}]{bogdanovic:2007.spin.alignment.shareddisk.merger}
{Bogdanovi{\'c}}, T., {Reynolds}, C.~S., \& {Miller}, M.~C. 2007, \apjl, 661,
  L147

\bibitem[{{Cattaneo} {et~al.}(2005){Cattaneo}, {Combes}, {Colombi}, {Bertin},
  \& {Melchior}}]{cattaneo:2005.mgr.agn.obsc}
{Cattaneo}, A., {Combes}, F., {Colombi}, S., {Bertin}, E., \& {Melchior}, A.
  2005, \mnras, 359, 1237

\bibitem[{{Colpi} {et~al.}(2007){Colpi}, {Callegari}, {Dotti}, {Kazantzidis},
  \& {Mayer}}]{colpi:2007.binary.in.mgrs}
{Colpi}, M., {Callegari}, S., {Dotti}, M., {Kazantzidis}, S., \& {Mayer}, L.
  2007, in American Institute of Physics Conference Series, Vol. 924, The
  Multicolored Landscape of Compact Objects and Their Explosive Origins, ed.
  {T.~di Salvo, G.~L.~Israel, L.~Piersant L.~Burderi G.~Matt A.~Tornambe \&
  M.~T.~Menna}, 705--714

\bibitem[{{Cox} {et~al.}(2006){Cox}, {Dutta}, {Di Matteo}, {Hernquist},
  {Hopkins}, {Robertson}, \& {Springel}}]{cox:kinematics}
{Cox}, T.~J., {Dutta}, S.~N., {Di Matteo}, T., {Hernquist}, L., {Hopkins},
  P.~F., {Robertson}, B., \& {Springel}, V. 2006, \apj, 650, 791

\bibitem[{{Croton} {et~al.}(2006)}]{croton:sam}
{Croton}, D.~J., {et~al.} 2006, \mnras, 365, 11

\bibitem[{{Davies} \& {Hunter}(1997)}]{davies:1997.counterrot.disk.instab}
{Davies}, C.~L., \& {Hunter}, Jr., J.~H. 1997, \apj, 484, 79

\bibitem[{{Di Matteo} {et~al.}(2005){Di Matteo}, {Springel}, \&
  {Hernquist}}]{dimatteo:msigma}
{Di Matteo}, T., {Springel}, V., \& {Hernquist}, L. 2005, \nat, 433, 604

\bibitem[{{Dotti} {et~al.}(2009){Dotti}, {Ruszkowski}, {Paredi}, {Colpi},
  {Volonteri}, \& {Haardt}}]{dotti:bh.binary.inspiral}
{Dotti}, M., {Ruszkowski}, M., {Paredi}, L., {Colpi}, M., {Volonteri}, M., \&
  {Haardt}, F. 2009, \mnras, 396, 1640

\bibitem[{{Dotti} {et~al.}(2010){Dotti}, {Volonteri}, {Perego}, {Colpi},
  {Ruszkowski}, \& {Haardt}}]{dotti:2010.spin.evol.recoil.fx}
{Dotti}, M., {Volonteri}, M., {Perego}, A., {Colpi}, M., {Ruszkowski}, M., \&
  {Haardt}, F. 2010, \mnras, 402, 682

\bibitem[{{El-Zant} \& {Shlosman}(2003)}]{elzant:nested.bar.scales}
{El-Zant}, A.~A., \& {Shlosman}, I. 2003, \apjl, 595, L41

\bibitem[{{Elitzur} \& {Shlosman}(2006)}]{elitzur:torus.wind}
{Elitzur}, M., \& {Shlosman}, I. 2006, \apjl, 648, L101

\bibitem[{{Englmaier} \& {Shlosman}(2004)}]{englmaier:nested.bar.decoupling}
{Englmaier}, P., \& {Shlosman}, I. 2004, \apjl, 617, L115

\bibitem[{{Erwin} \& {Sparke}(1999)}]{erwin:1999.triple.bar.structure.obs}
{Erwin}, P., \& {Sparke}, L.~S. 1999, \apjl, 521, L37

\bibitem[{{Erwin} \& {Sparke}(2002)}]{erwin:double.bar.obs}
---. 2002, \aj, 124, 65

\bibitem[{{Escala}(2007)}]{escala:nuclear.gas.transport.to.msigma}
{Escala}, A. 2007, \apj, 671, 1264

\bibitem[{{Fanidakis} {et~al.}(2011){Fanidakis}, {Baugh}, {Benson}, {Bower},
  {Cole}, {Done}, \& {Frenk}}]{fanidakis:2011.bh.spin.sam}
{Fanidakis}, N., {Baugh}, C.~M., {Benson}, A.~J., {Bower}, R.~G., {Cole}, S.,
  {Done}, C., \& {Frenk}, C.~S. 2011, \mnras, 410, 53

\bibitem[{{Feoli} \& {Mancini}(2009)}]{feoli:bhfp.1}
{Feoli}, A., \& {Mancini}, L. 2009, \apj, 703, 1502

\bibitem[{{Ferrarese} \& {Merritt}(2000)}]{FM00}
{Ferrarese}, L., \& {Merritt}, D. 2000, \apjl, 539, L9

\bibitem[{{Friedli} \& {Martinet}(1993)}]{friedli:1993.bars.win.bars.bulges}
{Friedli}, D., \& {Martinet}, L. 1993, \aap, 277, 27

\bibitem[{{Friedli} {et~al.}(1996){Friedli}, {Wozniak}, {Rieke}, {Martinet}, \&
  {Bratschi}}]{friedli:1996.disk.obs.multiple.substruct}
{Friedli}, D., {Wozniak}, H., {Rieke}, M., {Martinet}, L., \& {Bratschi}, P.
  1996, \aaps, 118, 461

\bibitem[{{Fruscione} {et~al.}(2005){Fruscione}, {Greenhill}, {Filippenko},
  {Moran}, {Herrnstein}, \& {Galle}}]{fruscione:abs.by.warped.disk}
{Fruscione}, A., {Greenhill}, L.~J., {Filippenko}, A.~V., {Moran}, J.~M.,
  {Herrnstein}, J.~R., \& {Galle}, E. 2005, \apj, 624, 103

\bibitem[{{Gallimore} {et~al.}(2006){Gallimore}, {Axon}, {O'Dea}, {Baum}, \&
  {Pedlar}}]{gallimore:2006.agn.outflow.gal.alignment}
{Gallimore}, J.~F., {Axon}, D.~J., {O'Dea}, C.~P., {Baum}, S.~A., \& {Pedlar},
  A. 2006, \aj, 132, 546

\bibitem[{{Gebhardt} {et~al.}(2000)}]{Gebhardt00}
{Gebhardt}, K., {et~al.} 2000, \apjl, 539, L13

\bibitem[{{Greenhill} \& {Gwinn}(1997)}]{greenhill:1997.maser.twist}
{Greenhill}, L.~J., \& {Gwinn}, C.~R. 1997, Astrophysics and Space Science,
  248, 261

\bibitem[{{Greenhill} {et~al.}(2003)}]{greenhill:circinus.acc.disk}
{Greenhill}, L.~J., {et~al.} 2003, \apj, 590, 162

\bibitem[{{Hasan} \& {Norman}(1990)}]{hasan:1990.bar.cmc.chaos}
{Hasan}, H., \& {Norman}, C. 1990, \apj, 361, 69

\bibitem[{{Hayward} {et~al.}(2011{\natexlab{a}}){Hayward}, {Kere{\v s}},
  {Jonsson}, {Narayanan}, {Cox}, \& {Hernquist}}]{hayward:2011.smg.merger.rt}
{Hayward}, C.~C., {Kere{\v s}}, D., {Jonsson}, P., {Narayanan}, D., {Cox},
  T.~J., \& {Hernquist}, L. 2011{\natexlab{a}}, \apj, 743, 159

\bibitem[{{Hayward}
  {et~al.}(2011{\natexlab{b}})}]{hayward:arepo.gadget.mergers}
{Hayward}, C.~C., {et~al.} 2011{\natexlab{b}}, \mnras, in preparation

\bibitem[{Heller {et~al.}(2001)Heller, Shlosman, \&
  Englmaier}]{heller:secondary.bar.instability}
Heller, C., Shlosman, I., \& Englmaier, P. 2001, The Astrophysical Journal,
  553, 661

\bibitem[{{Heller} \& {Shlosman}(1996)}]{heller:1996.dynamics.nuclear.rings}
{Heller}, C.~H., \& {Shlosman}, I. 1996, \apj, 471, 143

\bibitem[{{Hernquist} \& {Barnes}(1991)}]{hernquist:kinematic.subsystems}
{Hernquist}, L., \& {Barnes}, J.~E. 1991, \nat, 354, 210

\bibitem[{{Hickox} {et~al.}(2009)}]{hickox:multiwavelength.agn}
{Hickox}, R.~C., {et~al.} 2009, \apj, 696, 891

\bibitem[{{Hoffman} {et~al.}(2010){Hoffman}, {Cox}, {Dutta}, \&
  {Hernquist}}]{hoffman:mgr.orbit.structure.vs.fgas}
{Hoffman}, L., {Cox}, T.~J., {Dutta}, S., \& {Hernquist}, L. 2010, \apj, in
  press, arXiv:1001.0799

\bibitem[{{Hopkins}(2010)}]{hopkins:slow.modes}
{Hopkins}, P.~F. 2010, \mnras, in press, arXiv:1009.4702 [astro-ph]

\bibitem[{{Hopkins} {et~al.}(2009){Hopkins}, {Cox}, {Younger}, \&
  {Hernquist}}]{hopkins:disk.survival}
{Hopkins}, P.~F., {Cox}, T.~J., {Younger}, J.~D., \& {Hernquist}, L. 2009,
  \apj, 691, 1168

\bibitem[{{Hopkins} \& {Elvis}(2010)}]{hopkins:twostage.feedback}
{Hopkins}, P.~F., \& {Elvis}, M. 2010, \mnras, 401, 7

\bibitem[{{Hopkins} {et~al.}(2011{\natexlab{a}}){Hopkins}, {Hayward},
  {Narayanan}, \& {Hernquist}}]{hopkins:torus}
{Hopkins}, P.~F., {Hayward}, C.~C., {Narayanan}, D., \& {Hernquist}, L.
  2011{\natexlab{a}}, \mnras, in press, arXiv:1108.3086 [astro-ph]

\bibitem[{{Hopkins} \& {Hernquist}(2006)}]{hopkins:seyferts}
{Hopkins}, P.~F., \& {Hernquist}, L. 2006, \apjs, 166, 1

\bibitem[{{Hopkins} \& {Hernquist}(2009)}]{hopkins:mdot.dist}
---. 2009, \apj, 698, 1550

\bibitem[{{Hopkins} {et~al.}(2005{\natexlab{a}}){Hopkins}, {Hernquist}, {Cox},
  {Di Matteo}, {Martini}, {Robertson}, \&
  {Springel}}]{hopkins:lifetimes.methods}
{Hopkins}, P.~F., {Hernquist}, L., {Cox}, T.~J., {Di Matteo}, T., {Martini},
  P., {Robertson}, B., \& {Springel}, V. 2005{\natexlab{a}}, \apj, 630, 705

\bibitem[{{Hopkins} {et~al.}(2005{\natexlab{b}}){Hopkins}, {Hernquist}, {Cox},
  {Di Matteo}, {Robertson}, \& {Springel}}]{hopkins:lifetimes.obscuration}
{Hopkins}, P.~F., {Hernquist}, L., {Cox}, T.~J., {Di Matteo}, T., {Robertson},
  B., \& {Springel}, V. 2005{\natexlab{b}}, \apj, 632, 81

\bibitem[{{Hopkins} {et~al.}(2006{\natexlab{a}}){Hopkins}, {Hernquist}, {Cox},
  {Di Matteo}, {Robertson}, \& {Springel}}]{hopkins:qso.all}
---. 2006{\natexlab{a}}, \apjs, 163, 1

\bibitem[{{Hopkins} {et~al.}(2008){Hopkins}, {Hernquist}, {Cox}, \& {Kere{\v
  s}}}]{hopkins:groups.qso}
{Hopkins}, P.~F., {Hernquist}, L., {Cox}, T.~J., \& {Kere{\v s}}, D. 2008,
  \apjs, 175, 356

\bibitem[{{Hopkins} {et~al.}(2007{\natexlab{a}}){Hopkins}, {Hernquist}, {Cox},
  {Robertson}, \& {Krause}}]{hopkins:bhfp.theory}
{Hopkins}, P.~F., {Hernquist}, L., {Cox}, T.~J., {Robertson}, B., \& {Krause},
  E. 2007{\natexlab{a}}, \apj, 669, 45

\bibitem[{{Hopkins} {et~al.}(2007{\natexlab{b}}){Hopkins}, {Hernquist}, {Cox},
  {Robertson}, \& {Krause}}]{hopkins:bhfp.obs}
---. 2007{\natexlab{b}}, \apj, 669, 67

\bibitem[{{Hopkins} {et~al.}(2006{\natexlab{b}}){Hopkins}, {Narayan}, \&
  {Hernquist}}]{hopkins:old.age}
{Hopkins}, P.~F., {Narayan}, R., \& {Hernquist}, L. 2006{\natexlab{b}}, \apj,
  643, 641

\bibitem[{{Hopkins} \& {Quataert}(2010{\natexlab{a}})}]{hopkins:zoom.sims}
{Hopkins}, P.~F., \& {Quataert}, E. 2010{\natexlab{a}}, \mnras, 407, 1529

\bibitem[{{Hopkins} \& {Quataert}(2010{\natexlab{b}})}]{hopkins:m31.disk}
---. 2010{\natexlab{b}}, \mnras, 405, L41

\bibitem[{{Hopkins} \&
  {Quataert}(2011{\natexlab{a}})}]{hopkins:inflow.analytics}
---. 2011{\natexlab{a}}, \mnras, 415, 1027

\bibitem[{{Hopkins} \& {Quataert}(2011{\natexlab{b}})}]{hopkins:cusp.slopes}
---. 2011{\natexlab{b}}, \mnras, 411, L61

\bibitem[{{Hopkins} {et~al.}(2011{\natexlab{b}}){Hopkins}, {Quataert}, \&
  {Murray}}]{hopkins:rad.pressure.sf.fb}
{Hopkins}, P.~F., {Quataert}, E., \& {Murray}, N. 2011{\natexlab{b}}, \mnras,
  417, 950

\bibitem[{{Hopkins} {et~al.}(2011{\natexlab{c}}){Hopkins}, {Quataert}, \&
  {Murray}}]{hopkins:stellar.fb.winds}
---. 2011{\natexlab{c}}, \mnras, in press, arXiv:1110.4638 [astro-ph]

\bibitem[{{Hopkins} {et~al.}(2012){Hopkins}, {Quataert}, \&
  {Murray}}]{hopkins:fb.ism.prop}
---. 2012, \mnras, 421, 3488

\bibitem[{{Jacobs} \& {Sellwood}(2001)}]{jacobs:longlived.lopsided.disk.modes}
{Jacobs}, V., \& {Sellwood}, J.~A. 2001, \apjl, 555, L25

\bibitem[{{Kawakatu} \& {Wada}(2008)}]{kawakatu:disk.bhar.model}
{Kawakatu}, N., \& {Wada}, K. 2008, \apj, 681, 73

\bibitem[{{Keel}(1980)}]{keel:1980.seyfert.vs.galaxy.inclination}
{Keel}, W.~C. 1980, \aj, 85, 198

\bibitem[{{Kennicutt}(1998)}]{kennicutt98}
{Kennicutt}, Jr., R.~C. 1998, \apj, 498, 541

\bibitem[{{Keres} {et~al.}(2011){Keres}, {Vogelsberger}, {Sijacki}, {Springel},
  \& {Hernquist}}]{keres:2011.arepo.gadget.disk.angmom}
{Keres}, D., {Vogelsberger}, M., {Sijacki}, D., {Springel}, V., \& {Hernquist},
  L. 2011, \mnras, in press, arXiv:1109.4638

\bibitem[{{Kesden} {et~al.}(2010){Kesden}, {Sperhake}, \&
  {Berti}}]{kesden:2010.relativistic.kick.suppression}
{Kesden}, M., {Sperhake}, U., \& {Berti}, E. 2010, \apj, 715, 1006

\bibitem[{{King}(2003)}]{king:msigma.superfb.1}
{King}, A. 2003, \apjl, 596, L27

\bibitem[{{King}(2005)}]{king:msigma.superfb.2}
---. 2005, \apjl, 635, L121

\bibitem[{{King} {et~al.}(2005){King}, {Lubow}, {Ogilvie}, \&
  {Pringle}}]{king:2005.bh.acc.alignment.criterion}
{King}, A.~R., {Lubow}, S.~H., {Ogilvie}, G.~I., \& {Pringle}, J.~E. 2005,
  \mnras, 363, 49

\bibitem[{{King} \& {Pringle}(2006)}]{king:2006.chaotic.acc.lowspins}
{King}, A.~R., \& {Pringle}, J.~E. 2006, \mnras, 373, L90

\bibitem[{{King} \& {Pringle}(2007)}]{king:2007.align.radio.torus.notgal}
---. 2007, \mnras, 377, L25

\bibitem[{{King} {et~al.}(2008){King}, {Pringle}, \&
  {Hofmann}}]{king:2008.fragmentation.chaotic.accretion}
{King}, A.~R., {Pringle}, J.~E., \& {Hofmann}, J.~A. 2008, \mnras, 385, 1621

\bibitem[{{Kinney} {et~al.}(2000){Kinney}, {Schmitt}, {Clarke}, {Pringle},
  {Ulvestad}, \& {Antonucci}}]{kinney:2000.bh.jet.directions}
{Kinney}, A.~L., {Schmitt}, H.~R., {Clarke}, C.~J., {Pringle}, J.~E.,
  {Ulvestad}, J.~S., \& {Antonucci}, R.~R.~J. 2000, \apj, 537, 152

\bibitem[{{Kitsionas}
  {et~al.}(2009)}]{kitsionas:2009.grid.sph.compare.turbulence}
{Kitsionas}, S., {et~al.} 2009, \aap, 508, 541

\bibitem[{{Kollmeier} {et~al.}(2006)}]{kollmeier:eddington.ratios}
{Kollmeier}, J.~A., {et~al.} 2006, \apj, 648, 128

\bibitem[{{Kondratko} {et~al.}(2005){Kondratko}, {Greenhill}, \&
  {Moran}}]{kondratko:3079.acc.disk.maser}
{Kondratko}, P.~T., {Greenhill}, L.~J., \& {Moran}, J.~M. 2005, \apj, 618, 618

\bibitem[{{Kormendy} \& {Richstone}(1995)}]{KormendyRichstone95}
{Kormendy}, J., \& {Richstone}, D. 1995, \araa, 33, 581

\bibitem[{{Kumar} \&
  {Pringle}(1985)}]{kumar:1985.misalignment.bh.spin.viscous.disk}
{Kumar}, S., \& {Pringle}, J.~E. 1985, \mnras, 213, 435

\bibitem[{{Lagos} {et~al.}(2011){Lagos}, {Padilla}, {Strauss}, {Cora}, \&
  {Hao}}]{lagos:2011.agn.gal.orientation}
{Lagos}, C.~D.~P., {Padilla}, N.~D., {Strauss}, M.~A., {Cora}, S.~A., \& {Hao},
  L. 2011, \mnras, 414, 2148

\bibitem[{Laine {et~al.}(2002)Laine, Shlosman, Knapen, \&
  Peletier}]{laine:nested.bars.in.seyferts}
Laine, S., Shlosman, I., Knapen, J.~H., \& Peletier, R.~F. 2002, The
  Astrophysical Journal, 567, 97

\bibitem[{{Lauer} {et~al.}(1993)}]{lauer93}
{Lauer}, T.~R., {et~al.} 1993, \aj, 106, 1436

\bibitem[{{Lauer} {et~al.}(1996)}]{lauer:ngc4486b}
---. 1996, \apjl, 471, L79+

\bibitem[{{Lawrence}(1991)}]{lawrence:receding.torus}
{Lawrence}, A. 1991, \mnras, 252, 586

\bibitem[{{Lawrence} \& {Elvis}(1982)}]{lawrence:1982.torus.alignment}
{Lawrence}, A., \& {Elvis}, M. 1982, \apj, 256, 410

\bibitem[{{Levine} {et~al.}(2010){Levine}, {Gnedin}, \&
  {Hamilton}}]{levine:sim.mdot.pwrspectrum}
{Levine}, R., {Gnedin}, N.~Y., \& {Hamilton}, A.~J.~S. 2010, \apj, 716, 1386

\bibitem[{{Levine} {et~al.}(2008){Levine}, {Gnedin}, {Hamilton}, \&
  {Kravtsov}}]{levine2008:nuclear.zoom}
{Levine}, R., {Gnedin}, N.~Y., {Hamilton}, A.~J.~S., \& {Kravtsov}, A.~V. 2008,
  \apj, 678, 154

\bibitem[{{Livio} {et~al.}(1999){Livio}, {Ogilvie}, \&
  {Pringle}}]{livio:1999.non.spin.jet.from.accdisk}
{Livio}, M., {Ogilvie}, G.~I., \& {Pringle}, J.~E. 1999, \apj, 512, 100

\bibitem[{{Lodato} \& {Pringle}(2006)}]{lodato:2006.misaligned.acc.bh.spin}
{Lodato}, G., \& {Pringle}, J.~E. 2006, \mnras, 368, 1196

\bibitem[{{Maciejewski} \&
  {Athanassoula}(2008)}]{maciejewski:nested.bar.models}
{Maciejewski}, W., \& {Athanassoula}, E. 2008, \mnras, 389, 545

\bibitem[{{Maciejewski} \& {Sparke}(2000)}]{maciejewski:2000.bar.bar.orbits}
{Maciejewski}, W., \& {Sparke}, L.~S. 2000, \mnras, 313, 745

\bibitem[{{Magorrian} {et~al.}(1998)}]{magorrian}
{Magorrian}, J., {et~al.} 1998, \aj, 115, 2285

\bibitem[{{Mayer} {et~al.}(2007){Mayer}, {Kazantzidis}, {Madau}, {Colpi},
  {Quinn}, \& {Wadsley}}]{mayer:bh.binary.sph.zoom.sim}
{Mayer}, L., {Kazantzidis}, S., {Madau}, P., {Colpi}, M., {Quinn}, T., \&
  {Wadsley}, J. 2007, Science, 316, 1874

\bibitem[{{Moderski} \& {Sikora}(1996)}]{moderski:1996.chaotic.acc.lowspin}
{Moderski}, R., \& {Sikora}, M. 1996, \aaps, 120, C591

\bibitem[{{Natarajan} \& {Pringle}(1998)}]{natarajan:1998.jet.angle.vs.bh.spin}
{Natarajan}, P., \& {Pringle}, J.~E. 1998, \apjl, 506, L97

\bibitem[{{Nayakshin}(2005)}]{nayakshin:2005.warped.disk.obsc}
{Nayakshin}, S. 2005, \mnras, 359, 545

\bibitem[{{Nayakshin} \&
  {King}(2007)}]{nayakshin:forced.stochastic.accretion.model}
{Nayakshin}, S., \& {King}, A. 2007, \mnras, in press, arXiv:0705.1686

\bibitem[{{Nixon} {et~al.}(2011{\natexlab{a}}){Nixon}, {Cossins}, {King}, \&
  {Pringle}}]{nixon:2011.align.binary.fx.bh.mgrs}
{Nixon}, C.~J., {Cossins}, P.~J., {King}, A.~R., \& {Pringle}, J.~E.
  2011{\natexlab{a}}, \mnras, 412, 1591

\bibitem[{{Nixon} {et~al.}(2011{\natexlab{b}}){Nixon}, {King}, \&
  {Pringle}}]{nixon:2011.align.binary.acc.disk}
{Nixon}, C.~J., {King}, A.~R., \& {Pringle}, J.~E. 2011{\natexlab{b}}, \mnras,
  417, L66

\bibitem[{{Perego} {et~al.}(2009){Perego}, {Dotti}, {Colpi}, \&
  {Volonteri}}]{perego:2009.spin.evol.accretion}
{Perego}, A., {Dotti}, M., {Colpi}, M., \& {Volonteri}, M. 2009, \mnras, 399,
  2249

\bibitem[{{Pfenniger}(1984)}]{pfenniger:bar.dynamics}
{Pfenniger}, D. 1984, \aap, 134, 373

\bibitem[{{Pfenniger} \&
  {Norman}(1990)}]{pfenniger:1990.bar.buckling.dissipation}
{Pfenniger}, D., \& {Norman}, C. 1990, \apj, 363, 391

\bibitem[{{Price} \&
  {Federrath}(2010)}]{price:2010.grid.sph.compare.turbulence}
{Price}, D.~J., \& {Federrath}, C. 2010, \mnras, 406, 1659

\bibitem[{{Rigby} {et~al.}(2006){Rigby}, {Rieke}, {Donley}, {Alonso-Herrero},
  \& {P{\'e}rez-Gonz{\'a}lez}}]{rigby:qso.hosts}
{Rigby}, J.~R., {Rieke}, G.~H., {Donley}, J.~L., {Alonso-Herrero}, A., \&
  {P{\'e}rez-Gonz{\'a}lez}, P.~G. 2006, \apj, 645, 115

\bibitem[{{Risaliti} {et~al.}(1999){Risaliti}, {Maiolino}, \&
  {Salvati}}]{risaliti:seyfert.2.nh.distrib}
{Risaliti}, G., {Maiolino}, R., \& {Salvati}, M. 1999, \apj, 522, 157

\bibitem[{{Robertson} {et~al.}(2006){Robertson}, {Hernquist}, {Cox}, {Di
  Matteo}, {Hopkins}, {Martini}, \& {Springel}}]{robertson:msigma.evolution}
{Robertson}, B., {Hernquist}, L., {Cox}, T.~J., {Di Matteo}, T., {Hopkins},
  P.~F., {Martini}, P., \& {Springel}, V. 2006, \apj, 641, 90

\bibitem[{{Salow} \& {Statler}(2001)}]{salow:nuclear.disk.models}
{Salow}, R.~M., \& {Statler}, T.~S. 2001, \apjl, 551, L49

\bibitem[{{Salucci} {et~al.}(1999){Salucci}, {Szuszkiewicz}, {Monaco}, \&
  {Danese}}]{salucci:bhmf}
{Salucci}, P., {Szuszkiewicz}, E., {Monaco}, P., \& {Danese}, L. 1999, \mnras,
  307, 637

\bibitem[{{Sambhus} \& {Sridhar}(2002)}]{sambhus:m31.nuclear.disk.model}
{Sambhus}, N., \& {Sridhar}, S. 2002, \aap, 388, 766

\bibitem[{{Sanders} {et~al.}(1989){Sanders}, {Phinney}, {Neugebauer}, {Soifer},
  \& {Matthews}}]{sanders:1989.warped.qso.disk.ir.emission}
{Sanders}, D.~B., {Phinney}, E.~S., {Neugebauer}, G., {Soifer}, B.~T., \&
  {Matthews}, K. 1989, \apj, 347, 29

\bibitem[{{Schartmann} {et~al.}(2009){Schartmann}, {Meisenheimer}, {Klahr},
  {Camenzind}, {Wolf}, \&
  {Henning}}]{schartmann:2009.stellar.fb.effects.on.torus}
{Schartmann}, M., {Meisenheimer}, K., {Klahr}, H., {Camenzind}, M., {Wolf}, S.,
  \& {Henning}, T. 2009, \mnras, 393, 759

\bibitem[{{Scheuer} \& {Feiler}(1996)}]{scheuer:1996.bh.acc.disk.alignment}
{Scheuer}, P.~A.~G., \& {Feiler}, R. 1996, \mnras, 282, 291

\bibitem[{{Schmitt} {et~al.}(1997){Schmitt}, {Kinney}, {Storchi-Bergmann}, \&
  {Antonucci}}]{schmitt:1997.radio.alignment.w.host}
{Schmitt}, H.~R., {Kinney}, A.~L., {Storchi-Bergmann}, T., \& {Antonucci}, R.
  1997, \apj, 477, 623

\bibitem[{{Sellwood} \& {Merritt}(1994)}]{sellwood:1994.counterrot.bend.instab}
{Sellwood}, J.~A., \& {Merritt}, D. 1994, \apj, 425, 530

\bibitem[{{Shankar} {et~al.}(2004){Shankar}, {Salucci}, {Granato}, {De Zotti},
  \& {Danese}}]{shankar:bhmf}
{Shankar}, F., {Salucci}, P., {Granato}, G.~L., {De Zotti}, G., \& {Danese}, L.
  2004, \mnras, 354, 1020

\bibitem[{{Shaw} {et~al.}(1995){Shaw}, {Axon}, {Probst}, \&
  {Gatley}}]{shaw:1995.nuclear.bar.inflow}
{Shaw}, M., {Axon}, D., {Probst}, R., \& {Gatley}, I. 1995, \mnras, 274, 369

\bibitem[{{Shen} {et~al.}(2010){Shen}, {Shao}, \&
  {Gu}}]{shen:2010.torus.alignment}
{Shen}, S., {Shao}, Z., \& {Gu}, M. 2010, \apjl, 725, L210

\bibitem[{{Shlosman} \& {Heller}(2002)}]{shlosman:nested.bar.evol}
{Shlosman}, I., \& {Heller}, C.~H. 2002, \apj, 565, 921

\bibitem[{{Sijacki} {et~al.}(2011){Sijacki}, {Vogelsberger}, {Keres},
  {Springel}, \& {Hernquist}}]{sijacki:2011.gadget.arepo.hydro.tests}
{Sijacki}, D., {Vogelsberger}, M., {Keres}, D., {Springel}, V., \& {Hernquist},
  L. 2011, \mnras, in press, arXiv:1109.3468

\bibitem[{{Silk} \& {Rees}(1998)}]{silkrees:msigma}
{Silk}, J., \& {Rees}, M.~J. 1998, \aap, 331, L1

\bibitem[{{Simcoe} {et~al.}(1997){Simcoe}, {McLeod}, {Schachter}, \&
  {Elvis}}]{simcoe:1997.agn.host.alignment}
{Simcoe}, R., {McLeod}, K.~K., {Schachter}, J., \& {Elvis}, M. 1997, \apj, 489,
  615

\bibitem[{{Simpson} {et~al.}(1999){Simpson}, {Rawlings}, \&
  {Lacy}}]{simpson99:thermal.imaging.of.radio.gal}
{Simpson}, C., {Rawlings}, S., \& {Lacy}, M. 1999, \mnras, 306, 828

\bibitem[{{Soltan}(1982)}]{soltan82}
{Soltan}, A. 1982, \mnras, 200, 115

\bibitem[{{Springel}(2005)}]{springel:gadget}
{Springel}, V. 2005, \mnras, 364, 1105

\bibitem[{Springel(2010)}]{springel:arepo}
Springel, V. 2010, \mnras, 401, 791

\bibitem[{{Springel} {et~al.}(2005){Springel}, {Di Matteo}, \&
  {Hernquist}}]{springel:red.galaxies}
{Springel}, V., {Di Matteo}, T., \& {Hernquist}, L. 2005, \apjl, 620, L79

\bibitem[{{Springel} \& {Hernquist}(2003)}]{springel:multiphase}
{Springel}, V., \& {Hernquist}, L. 2003, \mnras, 339, 289

\bibitem[{{Tchekhovskoy} {et~al.}(2009){Tchekhovskoy}, {Narayan}, \&
  {McKinney}}]{tchekhovskoy:bh.spin.vs.radio.pwr}
{Tchekhovskoy}, A., {Narayan}, R., \& {McKinney}, J.~C. 2009, \apj,in press,
  arXiv:0911.2228

\bibitem[{{Tohline} \&
  {Durisen}(1982)}]{tohline:1982.gas.disk.orientation.in.prolate.gal}
{Tohline}, J.~E., \& {Durisen}, R.~H. 1982, \apj, 257, 94

\bibitem[{{Trump} {et~al.}(2009){Trump}, {Impey}, {Kelly}, {Elvis}, {Merloni},
  {Bongiorno}, {Gabor}, {Hao}, {McCarthy}, {Huchra}, {Brusa}, {Cappelluti},
  {Koekemoer}, {Nagao}, {Salvato}, \&
  {Scoville}}]{trump:2009.type1.agn.mdot.limits}
{Trump}, J.~R., {et~al.} 2009, \apj, in press [arXiv:0905.1123]

\bibitem[{{Ulvestad} \&
  {Wilson}(1984)}]{ulvestad:1984.radiojet.seyfert.misalignment}
{Ulvestad}, J.~S., \& {Wilson}, A.~S. 1984, \apj, 285, 439

\bibitem[{{van Meter} {et~al.}(2010){van Meter}, {Miller}, {Baker}, {Boggs}, \&
  {Kelly}}]{van-meter:2010.recoil.kick.formula}
{van Meter}, J.~R., {Miller}, M.~C., {Baker}, J.~G., {Boggs}, W.~D., \&
  {Kelly}, B.~J. 2010, \apj, 719, 1427

\bibitem[{{Vogelsberger} {et~al.}(2011){Vogelsberger}, {Sijacki}, {Keres},
  {Springel}, \& {Hernquist}}]{vogelsberger:2011.arepo.vs.gadget.cosmo}
{Vogelsberger}, M., {Sijacki}, D., {Keres}, D., {Springel}, V., \& {Hernquist},
  L. 2011, \mnras, in press arXiv:1109.1281

\bibitem[{{Volonteri} {et~al.}(2005){Volonteri}, {Madau}, {Quataert}, \&
  {Rees}}]{volonteri:2005.bh.spin.pred}
{Volonteri}, M., {Madau}, P., {Quataert}, E., \& {Rees}, M.~J. 2005, \apj, 620,
  69

\bibitem[{{Volonteri} \& {Rees}(2005)}]{volonteri:2005.bh.spin.sam}
{Volonteri}, M., \& {Rees}, M.~J. 2005, \apj, 633, 624

\bibitem[{{Wada} \& {Norman}(2002)}]{wada:starburst.torus.model}
{Wada}, K., \& {Norman}, C.~A. 2002, \apjl, 566, L21

\bibitem[{{Wada} {et~al.}(2009){Wada}, {Papadopoulos}, \&
  {Spaans}}]{wada:torus.mol.gas.hydro.sims}
{Wada}, K., {Papadopoulos}, P., \& {Spaans}, M. 2009, \apj, in press,
  arXiv:0906.5444

\bibitem[{{Wang} {et~al.}(2009){Wang}, {Hu}, {Li}, {Chen}, {King}, {Marconi},
  {Ho}, {Yan}, {Staubert}, \& {Zhang}}]{wang:2009.rad.eff.vs.z}
{Wang}, J.-M., {et~al.} 2009, \apjl, 697, L141

\bibitem[{{Willott} {et~al.}(2000){Willott}, {Rawlings}, {Blundell}, \&
  {Lacy}}]{willott00:optical.qso.frac.vs.l}
{Willott}, C.~J., {Rawlings}, S., {Blundell}, K.~M., \& {Lacy}, M. 2000,
  \mnras, 316, 449

\bibitem[{{Younger} {et~al.}(2008){Younger}, {Hopkins}, {Cox}, \&
  {Hernquist}}]{younger:minor.mergers}
{Younger}, J.~D., {Hopkins}, P.~F., {Cox}, T.~J., \& {Hernquist}, L. 2008,
  \apj, 686, 815

\bibitem[{{Zakamska} {et~al.}(2006)}]{zakamska:qso.hosts}
{Zakamska}, N.~L., {et~al.} 2006, \aj, 132, 1496

\bibitem[{{Zhang} {et~al.}(2009){Zhang}, {Soria}, {Zhang}, {Swartz}, \&
  {Liu}}]{zhang:2009.agn.vs.hubble.type}
{Zhang}, W.~M., {Soria}, R., {Zhang}, S.~N., {Swartz}, D.~A., \& {Liu}, J.~F.
  2009, \apj, 699, 281

\end{thebibliography}

\end{document}